\documentclass[reprint,amssymb,amsmath,aps,prl,nofootinbib]{revtex4-1}

\usepackage{graphicx}
\usepackage{dcolumn}
\usepackage{bm}
\usepackage{hyperref}
\usepackage{amsthm}
\usepackage{relsize}

\usepackage[abs]{overpic}

\usepackage{xcolor,varwidth}

\newtheorem*{theorem}{Theorem}

\newcommand{\la}{\langle}
\newcommand{\ra}{\rangle}

\newcommand{\GS}{{\rm GS}}

\newcommand{\Htot}{\hat H}

\newcommand{\Hh}{\hat H_{\rm f}}
\newcommand{\Eh}{E_{\rm f}}

\newcommand{\Hi}{\hat H_{\rm i}}

\newcommand{\U}{\hat U}

\newcommand{\rr}{{\mathbf r}}

\newcommand{\vv}{{\mathbf v}}

\newcommand{\q}{{\mathbf q}}

\newcommand{\ppi}{\hat {\mathbf P}_{\rm i}}

\newcommand{\disp}{{\cal E}}

\newcommand{\be}{\begin{equation}}
\newcommand{\ee}{\end{equation}}

\newcommand{\ket}[1]{\left|{#1}\right\rangle}

\begin{document}


\title{Perpetual motion and driven dynamics \\ of a mobile impurity in a quantum fluid}

\author{O. Lychkovskiy}
\affiliation{ Russian Quantum Center, Novaya St. 100A, Skolkovo, Moscow Region, 143025, Russia.}


\begin{abstract}

We study the dynamics of a mobile impurity in a quantum fluid at zero temperature. Two related settings are considered. In the first setting the impurity is injected  in the fluid with some initial velocity $\vv_0$, and we are interested in its velocity at infinite time, $\vv_\infty$.  We derive a rigorous upper bound on $|\vv_0-\vv_\infty|$ for initial velocities smaller than the generalized critical velocity. In the limit of vanishing impurity-fluid coupling this bound amounts to $\vv_\infty=\vv_0$ which can be regarded as a rigorous proof of the Landau criterion of superfluidity. In the case of a finite coupling the velocity of the impurity can drop, but not to zero;  the bound  quantifies the maximal possible drop. In the second setting a small constant force is exerted upon the impurity. We argue that two distinct dynamical regimes exist -- backscattering oscillations of the impurity velocity and saturation of the velocity without oscillations. For fluids with $v_{c {\rm L}}=v_s$ (where  $v_{c {\rm L}}$ and $v_s$ are the Landau critical velocity and sound velocity, respectively) the latter regime is realized. For fluids with $v_{c {\rm L}} < v_s$ both regimes are possible. Which regime is realized in this case depends on the mass of the impurity, a nonequilibrium quantum phase transition occurring at some critical mass. Our results are equally valid in one, two and three dimensions.
\end{abstract}

\maketitle


{\em Introduction.}---
What happens to an impurity particle injected in a quantum fluid
at zero temperature?
According to the Landau criterion of superfluidity  \cite{landau1941JETP} generalized to account for motion of a particle of a finite mass \cite{rayfield1966roton}, if the initial velocity of the impurity $v_0$ is less than the (mass-dependent) generalized critical velocity $v_c$, the impurity keeps moving forever without dissipation.\footnote{A necessary condition for $v_c>0$ is that that the dispersion of the fluid is not identically zero, which we assume throughout the paper. This means that we consider two- and three-dimensional superfluids and generic one-dimensional fluids, but not e.g. Fermi liquids. In practical terms, our consideration can be relevant for superfluid helium and metastable quantum fluids realized in ultracold atom experiments.} However, the kinematical argument beyond the generalized Landau criterion \cite{landau1941JETP,rayfield1966roton} is nonrigorous: It is based on the assumption that the {\it kinetic} energy is conserved, which is valid only approximately. Generally speaking, this argument does not exclude the possibility that corrections to the above approximation build up with time in such a way that the velocity of the impurity does relax to a zero or nonzero value in the long run \cite{suzuki2014creation,roberts2005casimir,roberts2006force,roberts2009superfluids}. Indeed, numerical and semi-numerical calculations for specific systems has shown that the velocity does drop below $v_0$ even when $v_0<v_c$ \cite{mathy2012quantum,shashi2014radio}. As a rule, numerical calculations are limited to finite times and therefore can not unambiguously provide an infinite-time asymptotic value of the velocity, $v_\infty$. In particular, an important qualitative question -- whether the impurity eventually stops -- often remains unanswered.
This issue has been recently addressed in the context of a specific model: An  upper bound on $|v_0-v_\infty|$ has been rigorously derived for the impurity injected in the one-dimensional (1D) gas of free fermions \cite{Lychkovskiy2013}. The first goal of the present paper is to provide an analogous bound valid for an  arbitrary quantum fluid in any dimensionality. We rigorously prove that $|\vv_0-\vv_\infty|$ is bounded from above for $|\vv_0|<v_c$, the bound depending on the dispersion of the fluid, strength of the coupling between the impurity and the fluid,  mass of the impurity and its initial velocity. In the limit of vanishing impurity-fluid coupling the bound reduces to $\vv_\infty=\vv_0$, in accordance with the generalized Landau criterion of superfluidity \cite{landau1941JETP,rayfield1966roton}. In the case of finite interaction the bound  quantifies the maximal possible drop of the velocity.

The second question we address is as follows: What happens to an impurity immersed in a quantum fluid at zero temperature and pulled by a small constant force? This question was previously studied for impurities in superfluid helium \cite{bowley1975roton,bowley1977motion,allum1977breakdown} and, recently, in 1D fluids \cite{Gangardt2009,schecter2012dynamics,schecter2012critical,Gamayun2014,Gamayun2014keldysh}. It was found that impurities in helium exhibit sawtooth velocity oscillations emerging from backscattering on rotons \cite{bowley1975roton,bowley1977motion,allum1977breakdown}. Similar {\it backscattering oscillations} (BO) have been found in 1D Tonks-Girardeau gas but only for sufficiently heavy impurities \cite{Gamayun2014}. For lighter impurities another dynamical regime has been observed -- {\it saturation of the velocity without oscillations} (SwO). Basing on the  same kinematical constraint which underlies the generalized Landau argument \cite{landau1941JETP,rayfield1966roton}, we investigate how general quantum fluids can be classified with respect to the regimes of driven dynamics. We find that BO and SwO are the only two generic regimes. A criterion determining which one is realized for a particular fluid and impurity is derived.

It is worth emphasizing that all the methods and results presented in this Letter are universally valid both for one-dimensional fluids and higher-dimensional fluids despite the well-known dramatic difference between the former and the latter with respect to the structure of elementary excitations \cite{Giamarchi2003}. This constitutes the major advancement over recent works \cite{mathy2012quantum,knap2014quantum,Lychkovskiy2013,Gangardt2009,schecter2012dynamics,schecter2012critical,Gamayun2014,Gamayun2014keldysh,gamayun2015impurity} focused on 1D fluids which explicitly invoked special features of physics in one dimension.

\begin{figure}[t]
\includegraphics[width= \linewidth]{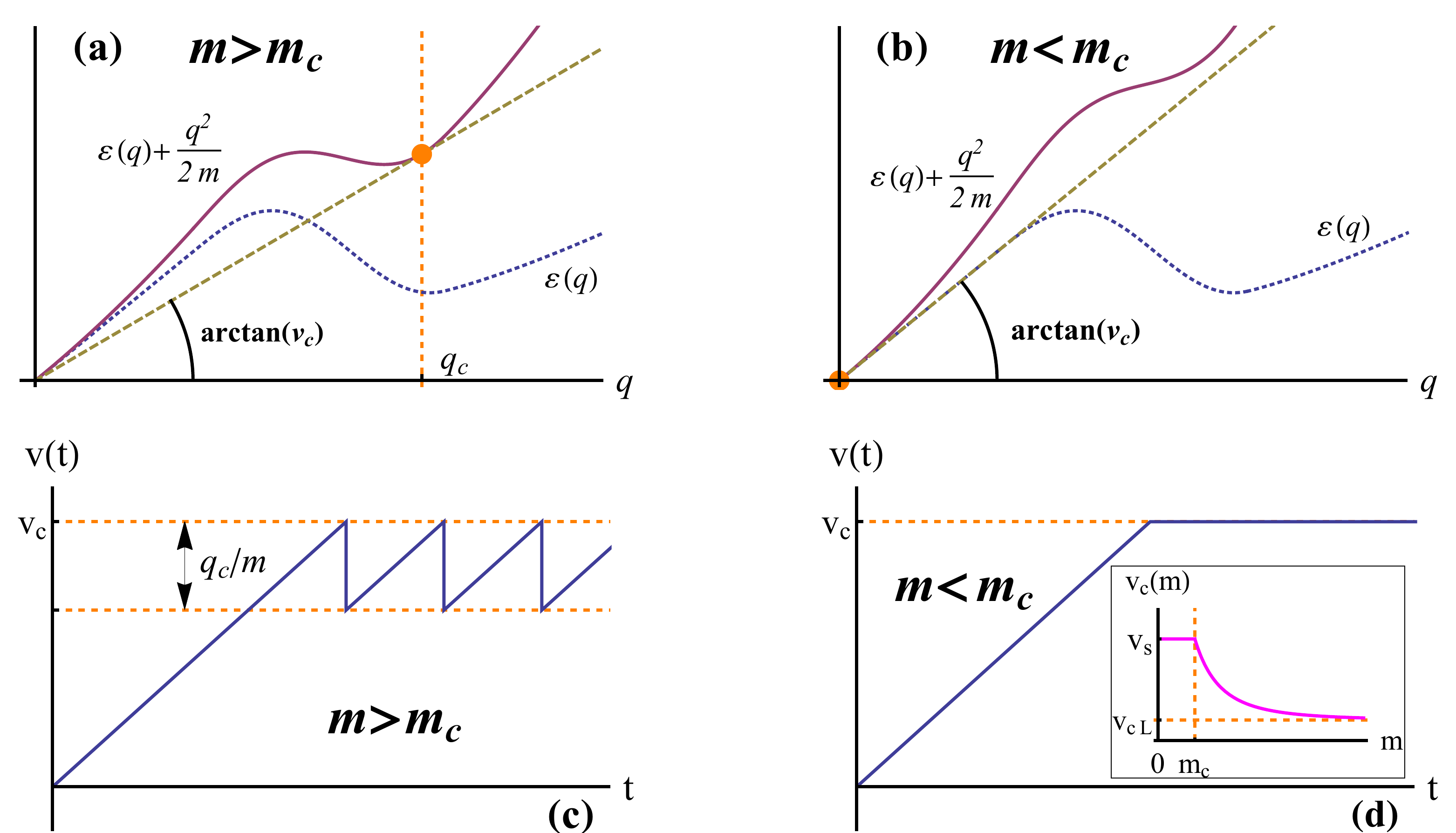}
\caption{\label{fig 1}(color online) (a) and (b): Geometrical illustration of definitions of generalized critical velocity $v_c$, eq. \eqref{critical velocity}, and critical momentum transfer $q_c$, eq. \eqref{backscattering momentum}. $v_c$ is smaller than the sound velocity $v_s$ for $m>m_c$ (a), while $v_c=v_s$ for $m<m_c$ (b). The thick dot (orange online) marks the position of the critical momentum transfer $q_c$ which is finite for $m>m_c$ but vanishes for $m<m_c$. (c) and (d): Velocity of the impurity pulled by a small constant force {\it vs.} time.
Backscattering oscillations occur for $m>m_c$ (c). Velocity of the impurity saturates at $v_c$ without oscillations for $m<m_c$ (d). Inset: Generalized critical velocity as a function of the impurity mass.
In the limit of $m\to\infty$ the generalized critical velocity approaches the Landau critical velocity  $v_{c \,{\rm L}}$.}
\end{figure}

{\em Setup and notations.}---
We consider a single impurity particle immersed in a quantum fluid.
The Hamiltonian of the combined impurity-fluid system reads
$
\Htot = \Hh+\Hi+\U,
$
where $\Hh$, $\Hi$ and $\U$ describe the fluid, the impurity and the impurity-fluid interaction, respectively.
$\Hh$, $\Hi$ and $\U$ are translationally invariant and isotropic (the latter requirement can be dropped at the price of the results and derivation being more bulky).
An eigenstate of $\Hh$ with an energy $\Eh$  is denoted by $\ket{\Eh}$. Each $\ket{\Eh}$ is also an eigenstate of  the momentum.
The dispersion of the fluid, $\varepsilon(q)$, is defined as
a minimal eigenenergy which corresponds to a given momentum $\q$ with $|\q|=q$.

We  use a special notation, $\ket{\GS}$, for the ground state of the fluid. We set the ground state energy of the fluid to zero and assume that the momentum in the ground state is zero.
This implies $\varepsilon(0)=0$ and $\varepsilon(q)\geq 0$.
The speed of sound is defined as
$
v_s\equiv\varepsilon'(0).
$
Note that we do not impose any restrictions on the strength of interactions between the elementary excitations of the fluid.

The Hamiltonian of the impurity reads
$
\Hi =\ppi^2/(2m),
$
where  $\ppi$ is the momentum of the impurity.
%
%
%
Interaction $\U$ is pairwise with an interaction potential $U(r)$.
We call the interaction {\it everywhere repulsive} whenever 
\be\label{potential}
U(r)\geq 0 ~~~~~~ \forall r.
\ee

We denote product eigenstates of $\Hh+\Hi$ by
$
\ket{\Eh,\vv} \equiv |\Eh \rangle \otimes |\vv\rangle,
$
where $|\vv \rangle$ is the plane wave of the impurity with the momentum $m \vv$. Initially the impurity-fluid system is in a product state
$
|\GS, \vv_0\rangle = |\GS \rangle \otimes |\vv\rangle,
$ i.e. the impurity moves in the  fluid at zero temperature with velocity $\vv_0$.\footnote{This initial state can be realized in ultracold atom experiments by accelerating a noninteracting impurity inside the atomic cloud and switching the impurity-atom interaction by means of the  Feshbach resonance afterwards.}
Since the total momentum is an integral of motion, in what follows we restrict all operators to the subspace with the total momentum $m\vv_0$.

The quantity we are interested in is the velocity of the impurity at infinite time. It is defined as
\be\label{vinfty definition}
\vv_\infty \equiv \frac{1}{m} \lim_{t\rightarrow\infty}\frac1t\int_0^t dt' \langle \GS, \vv_0|e^{i \hat H t'}\ppi e^{-i \hat H t'} | \GS, \vv_0 \rangle.
\ee

Expanding the initial state in eigenstates $\ket{E}$ of the total Hamiltonian, $\Htot$,  and integrating out oscillating exponents, one obtains
\be\label{vinfty}
\vv_\infty = \frac{1}{m}\sum_{\ket{E}}\big|\la \GS,\vv_0|E \ra\big|^2 \la E|\ppi|E\ra.
\ee
Note that if $\Htot$ has degenerate eigenvalues, one should adjust the eigenbasis to diagonalize the matrix $\la E'|\GS, \vv_0\ra\la\GS, \vv_0|E\ra$ in every degenerate subspace.

\begin{table}[t]
\begin{ruledtabular}
\begin{tabular}{lccc}
        & $v_{c {\rm L}}=v_s$   &  \multicolumn{2}{c}{$v_{c {\rm L}}<v_s$}\\
        &                       & $m<m_c$                       &  $m>m_c$  \\
regime & SwO                  & SwO                          & BO \\
\end{tabular}
\end{ruledtabular}
\caption{\label{table}%
Conditions determining which of the two dynamical regimes -- backscattering oscillations (BO) or saturation without oscillations (SwO) -- is realised in a specific fluid for a specific mass of the impurity.
}
\end{table}

{\em Perpetual motion.}---
We start from reviewing kinematical arguments which lead to the notion of critical  velocity \cite{landau1941JETP,rayfield1966roton}. Consider an impurity with a velocity $\vv_0$ which scatters off the fluid which is initially in its ground state. Assume that the impurity can not form a bound state with particles of the fluid.  Assume further that the final state of the impurity-fluid system is a product eigenstate of noninteracting Hamiltonian $\Hh+\Hi$, $\q$ and $\Eh\geq \varepsilon(q)$ being respectively final momentum and energy of the fluid. If one disregards the contribution of the impurity-fluid coupling to its energy, then conservation laws lead to
\be\label{energy conservation}
v_0 q \geq \vv_0 \q = \Eh + \frac{q^2}{2 m} \geq \varepsilon(q) + \frac{q^2}{2 m},
\ee
where $v_0 \equiv |\vv_0|$.
If $v_0$ is sufficiently small, $v_0<v_c$, then for all  $\q \neq 0$ the inequality \eqref{energy conservation} can not be fulfilled. The generalized critical velocity $v_c$ is defined as \cite{rayfield1966roton}
\be\label{critical velocity}
v_c \equiv \inf_{q} \frac{\varepsilon(q) + \frac{q^2}{2 m}}{q}.
\ee
Physically, $v_c$ is the minimal velocity which allows the impurity to create real excitations of the fluid, in the approximation of noninteracting final impurity-fluid state. The geometrical sense of the generalized critical velocity can be seen from Fig.~\ref{fig 1}: The line $v_c q$ is a tangent to the curve $\varepsilon(q) + \frac{q^2}{2 m}$.
Originally Landau defined the critical velocity in the limit $m\rightarrow \infty$ \cite{landau1941JETP}:
\be
v_{c {\rm L}} \equiv \inf_{q}\left( \varepsilon(q)/q \right).
\ee
Note that $v_{c {\rm L}}$ is an attribute of the fluid alone while $v_c$ is an attribute of the impurity-fluid system.

In is worth emphasizing that the definition of the generalized critical velocity \eqref{critical velocity}, although motivated by the Landau argument, stands alone and will be used beyond the scope of this argument in what follows.


Clearly, the argument by Landau reviewed above
is not rigorous: The impurity-fluid interaction is largely disregarded, its role being merely to justify why the transition from the initial to a final state occurs at all.
Our aim is to derive a rigorous relation between $\vv_0$ and $\vv_\infty$. To this end we prove the following
%
\begin{theorem}
Consider an impurity particle immersed in a quantum fluid. Initially the
system is prepared in the product state $|\GS, \vv_0\rangle$ with the initial velocity of the impurity
$
v_0 \equiv |\vv_0| < v_c.
$
The difference between the initial and  infinite-time velocities of the impurity is bounded from above according to
\be\label{lower bound general}
\begin{array}{ll}
|\vv_0 -\vv_\infty| \leq &   \frac{1}{m(v_c-v_0)}
\Big(
\langle \GS, \vv_0| \U | \GS, \vv_0\rangle-
 \\
&
\sum\limits_{\ket{E}}\big|\la \GS,\vv_0|E \ra\big|^2 \langle E| \U | E \rangle
\Big).
\end{array}
\ee
%
If the interaction between the impurity and the fluid is everywhere repulsive, i.e. the condition  (\ref{potential}) is fulfilled, then a more transparent bound holds:
\be\label{lower bound}
|\vv_0 -\vv_\infty| \leq \frac{  \overline U}{m(v_c-v_0)},
\ee
where $\overline U \equiv\int d \rr \,\rho \, U(|\rr|)$ and $\rho$ is the number density of the particles of the fluid.
\end{theorem}
This theorem generalises an analogous result obtained in \cite{Lychkovskiy2013} for a specific one-dimensional  fluid.  \\
{\em Proof.}
According to \eqref{vinfty}
\begin{align}\label{intermediate 1}
 &
|\vv_0-\vv_\infty| = \nonumber\\
 & =
\Big|
\sum_{\ket{E}}\sum_{\ket{\Eh,\vv}} (\vv_0-\vv)\big|\la E| \Eh,\vv \ra\big|^2 \big|\la \GS,\vv_0|E \ra\big|^2
\Big| \nonumber \\
 & \leq
\sum_{\ket{E}}\left(\sum_{\ket{\Eh,\vv}} |\vv_0-\vv| \big|\la E| \Eh,\vv \ra\big|^2 \right) \big|\la \GS,\vv_0|E \ra\big|^2
.
\end{align}
The sums are performed over the eigenstates $\ket{E}$ of  $\hat H$ and over the eigenstates   $\ket {\Eh,\vv}$ of ~$\Hh+\Hi$ with the total momentum $m\vv_0$.

The key step is to notice that according to \eqref{critical velocity}
\be\label{key inequality}
|\vv_0-\vv| \leq  \frac{1}{m(v_c-v_0)} (\Eh+\frac{m\vv^2}{2}-\frac{m \vv_0^2}{2})
\ee
for any $\ket {\Eh,\vv}$ with the total momentum $m \vv_0$.
This inequality is of pure kinematical origin.
It leads to
\begin{align}\label{intermediate 2}
 &\sum_{\ket{\Eh,\vv}} |\vv_0-\vv| \big|\la E| \Eh,\vv \ra\big|^2 \leq \nonumber\\
 &\leq
\frac{1}{m(v_c-v_0)} \sum_{\ket{\Eh,\vv}} \la E| \Hh+\Hi -\frac{\vv_0^2}{2m}|\Eh,\vv \ra \la  \Eh,\vv | E\ra \nonumber\\
&= \frac{1}{m(v_c-v_0)}\left(  E -\frac{\vv_0^2}{2m} - \la E| \U | E\ra\right).
\end{align}
Substituting eq. \eqref{intermediate 2} into   eq. \eqref{intermediate 1} one obtains the desired bound \eqref{lower bound general}.

If the impurity-fluid coupling is everywhere repulsive, one obtains the bound \eqref{lower bound} from the bound \eqref{lower bound general} by omitting  the second term in the brackets in the r.h.s. of \eqref{lower bound general} and rewriting the first term according to
$
\la \GS,\vv_0| \U | \GS,\vv_0\ra= \overline U
$.
$\blacksquare$

In the reminder of the present section we discuss the above theorem. First, we stress that the bounds \eqref{lower bound general} and \eqref{lower bound} hold for an arbitrary  interacting quantum fluid in  arbitrary dimensions, in contrast to an earlier result \cite{Lychkovskiy2013} valid for a one-dimensional gas of  free fermions.  Remarkably, interactions between elementary excitations of the fluid renormalize $\varepsilon(p)$ but do not enter the bounds explicitly. Moreover, $\varepsilon(p)$ itself enters the bounds only through $v_c$.

Consider implications of the theorem in the  weak impurity-fluid coupling limit. To define the latter we introduce a family of interaction potentials $U_\gamma(r)=\gamma U_1(r)$ parameterized by the dimensionless coupling $\gamma$. The weak coupling limit amounts to considering small $\gamma$ (i.e. expanding all quantities of interest around $\gamma=0$) after the thermodynamic limit ($N\to \infty$,  $V=N/\rho$, $\rho$ fixed) is taken.  The physical meaning of this limit is that the interaction energy is small compared to the total energy per particle but large compared to the level spacing.

The Landau criterion of superfluidity \cite{landau1941JETP} (generalised for impurities of finite mass \cite{rayfield1966roton}) can be rigorously  proved in the  weak  coupling limit by virtue of the bound \eqref{lower bound general}. To this end, if the interaction is not everywhere repulsive, we invoke an additional, rather natural assumption that $\langle E| \U_1 | E \rangle \geq -C$ for any $|E\rangle$, where $C \geq 0$ is some constant independent on $N$ and $|E\rangle$. For example, if bound states of the impurity particle and particles of the fluid exist, $C$ is expected to be of order of the largest binding energy among all such "molecules". The case of everywhere repulsive interaction amounts to $C=0$. The bound \eqref{lower bound general}
complemented by the aforementioned assumption immediately leads to the Landau's statement $\vv_\infty=\vv_0+O(\gamma)$ for $|\vv_0|<v_c$ in the weak coupling limit.

It is worth emphasising that a straightforward perturbation theory in $\gamma$ does not lead to a correct many-body overlap  $\big|\la \GS,\vv_0|E \ra\big|^2$  (see a thorough discussion of
this point in \cite{march1967many}), and, as a consequence, does not permit a universal calculation of $v_\infty$ directly from eq. \eqref{vinfty}.
This problem does not emerge when treating the r.h.s. of the bound \eqref{lower bound general} because the interaction term $\U$ enters the latter explicitly.

Since the bound \eqref{lower bound general} invokes exact many-body eigenstates, its immediate application beyond the perturbative regime is possible only for integrable systems. These include (i) an impurity in a 1D gas of free fermions or infinitely repulsive bosons \cite{mcguire1965interacting} and (ii) an impurity in a 1D gas of bosons, with masses of the impurity and host boson being equal, as well as boson-boson and boson-impurity couplings being equal (bosonic Yang-Gaudin model \cite{yang1967some, gaudin1983fonction}). In the former model it is possible to calculate $v_\infty$ directly by means of eq. \eqref{vinfty} \cite{Burovski2013} (see also \cite{gamayun2015impurity}). In the latter, more sophisticated model, an analogous analytical calculation would likely to be much more intricate (if ever possible) since calculating overlaps $\big|\la \GS,\vv_0|E \ra\big|^2$ within Bethe ansatz is a hard task. On the other hand, application of the bound \eqref{lower bound general} should be feasible in this model since it requires a much simpler calculation of a  matrix element of a local operator. In the nonintegrable cases the bound \eqref{lower bound general} should be supplemented by some approximate method for calculating $\langle E| \U | E \rangle$ (e.g. perturbation theory, as is exemplified by the proof of Landau criterion presented above).

Now we turn to the bound \eqref{lower bound}. Though valid for a narrower class of interactions, it has the advantage of simplicity compared to the bound  \eqref{lower bound general} and can be easily applied without resorting to any approximations and limits. An additional benefit of the bound \eqref{lower bound} is that it obviates two important points. First, the bound holds equally well for a finite fluid and in the thermodynamic limit. Second, the inequality (\ref{lower bound}) represents a nontrivial bound even for   long range interactions, provided the interaction potential decreases with distance faster than~$1/r^D$, $D$ being the dimensionality of the system. The latter requirement ensures that $\overline U$ does not diverge at large distances. We expect that both observation generically hold for the bound \eqref{lower bound general} as well.

Possible divergence of $\overline U$ deserves further discussion. It can also emerge at small $r$. In particular, it prevents us from considering hard sphere impurity-fluid interaction. Divergence in $\overline U$  implies that the initial state $|\GS, \vv_0\rangle$ has divergent energy and thus the problem is ill-formulated from the outset. How to correctly formulate the problem in this situation is an interesting open question.

We exemplify the usage of the bound  \eqref{lower bound} in one and three dimensions. In the case of one dimension, we consider the pointlike repulsive impurity-fluid potential $U(x)=(U_0/\rho)\delta(x)$ with $U_0>0$ to obtain $|\vv_0 -\vv_\infty|\leq  U_0  \left(m(v_c-v_0)\right)^{-1}$. In the context of ultracold atom experiments this potential is an excellent low-energy approximation to any real impurity-fluid coupling with positive scattering length $a$, $U_0$ being a function of $a$ and transverse confinement energy  \cite{olshanii1998atomic}. This result has been earlier obtained for a special case of an impurity in a 1D gas of free fermions \cite{Lychkovskiy2013}; here it is proven for an arbitrary interacting 1D fluid.

In the case of three dimensions, we consider a ``square'' potential $U(r)=U_0 \theta(r_0-r)$. In this case the bound reads $|\vv_0 -\vv_\infty|\leq (4\pi/3)r_0^3 \rho U_0  \left(m(v_c-v_0)\right)^{-1}$. In the limit when the interaction range $r_0$ is much larger than the scattering length $a\simeq 2 \mu U_0 r_0^3/(3\hbar^2)$ (with $\mu$ being reduced mass) the bound can be expressed through the scattering length: $|\vv_0 -\vv_\infty|\lesssim 2\pi \hbar^2 a \rho \left(m\mu(v_c-v_0)\right)^{-1}$.


It is instructive to compare the above theorem with a rigorous result obtained in \cite{knap2014quantum}: The expectation value of the impurity velocity in the momentum-dependent ground state  equals to the slope of the {\it total} dispersion of the impurity-fluid system which is generically nonzero.
Thus Ref. \cite{knap2014quantum} proves the very possibility of the perpetual motion of an impurity in a quantum fluid. However, it does not relate the initial velocity of the injected impurity, $v_0$, to its asymptotic velocity $v_\infty$, in contrast to the theorem presented above.

{\em Dynamics of driven impurity.}---
In the present section we consider an impurity weakly coupled to a fluid and driven by a small constant force.  The kinematical reasoning summarized in the beginning of the previous section can be extended to the case with driving. This was done for mobile impurities in superfluid helium in Refs. \cite{bowley1975roton,bowley1977motion,allum1977breakdown}. We study a problem in a wider context of an arbitrary quantum fluid.

Consider the impurity to be initially at rest.  The force accelerates it freely until its velocity reaches $v_c$. At this instant the impurity acquires a chance to scatter off the fluid. It is clear from eq. \eqref{energy conservation} that the scattering channel which opens first is the back scattering. In this process the impurity loses some  momentum $q_c$ which is transferred to the fluid. The critical momentum transfer $q_c$ delivers minimum in eq. \eqref{critical velocity}:
\be\label{backscattering momentum}
v_c q_c =  \varepsilon(q_c) + \frac{q_c^2}{2 m}.
\ee
The geometrical meaning of $q_c$ is illustrated in Fig. \ref{fig 1}~(a),(b): the line $v_c q$ touches the curve $\varepsilon(q) + \frac{q^2}{2 m}$ in the point $(q_c,v_c q_c)$. Note that $q_c$ is unrelated to $m v_c$.

Up to this point our presentation has closely followed Refs. \cite{bowley1975roton,bowley1977motion,allum1977breakdown}.
The central new observation is that the behavior of the impurity depends crucially on whether or not $q_c$ is zero. Consider first the case $q_c>0$ (see Fig \ref{fig 1}~(a)) which is relevant, in particular, for impurities in helium \cite{bowley1975roton,bowley1977motion,allum1977breakdown}. After the first scattering the velocity of the impurity drops by
$
\Delta v=q_c/m,
$
and the impurity starts to freely accelerate until its velocity again reaches $v_c$, after which the whole cycle is repeated. This is how backscattering oscillations emerge \cite{bowley1975roton,bowley1977motion,allum1977breakdown}.

Consider now the case when $q_c=0$, see Fig \ref{fig 1} (b). This case was not considered in \cite{bowley1975roton,bowley1977motion,allum1977breakdown} since it can not be realized with realistic impurities in superfluid helium (see below). In this case, as soon as the velocity of the impurity reaches $v_c$, the impurity starts to dissipate the pumped energy by producing infrared excitations of the fluid. In the limit of small force this leads to the   saturation of its velocity at  $v_c$  without oscillations (SwO).

One can see that whether or not $q_c$ is zero governs which of the two generic regimes, SwO or BO, is realized for a particular fluid and impurity. Note that $q_c=0$ ($q_c>0$) whenever $v_c=v_s$ ($v_c<v_s$), see Fig. \ref{fig 1}. The relations between $v_c$ and $v_s$, in turn, is determined by the Landau critical velocity of the fluid, $v_{c {\rm L}}$ and the mass of the impurity, $m$. As a result, in the fluid with $v_{c {\rm L}}=v_s$ (e.g. in the Bogolyubov gas of weakly coupled bosons) only SwO is possible, regardless of value of $m$.  In contrast, in the fluid with $v_{c {\rm L}}<v_s$  both  SwO and BO are possible, depending on the mass of the impurity: BO emerge in the case of a heavy impurity,  $m>m_c$,    while SwO takes place for a light impurity, $m<m_c$. The critical mass $m_c$ is determined from the  equation
$
v_c(m_c)=v_s,
$
in which we explicitly indicate the dependence of the generalized critical velocity on the mass of the impurity, see eq. \eqref{critical velocity} and the inset in Fig. \ref{fig 1}. The amplitude of BO generically experiences a jump from a finite value to zero at $m=m_c$. Thus if one regards $m$ as a tunable parameter, the transition over $m_c$ is a nonequilibrium quantum phase transition. Conditions discriminating between the two dynamical regimes are summarized in Table \ref{table}. 
Note that SWO was not observed in superfluid helium since sufficiently light impurities were lacking.

Existence of two dynamical regimes separated by a nonequilibrium quantum phase transition is consistent with the results of the detailed study of a specific 1D fluid \cite{Gamayun2014}.



BO get damped at finite forces since the direction (for $D>1$)  and the value (for any dimensionality) of the momentum transfer vary from one scattering event to another. In Ref. \cite{Gamayun2014} a kinetic theory for an impurity in the Tonks-Girardeau gas has been developed and the damping rate has been calculated. This theory can be generalized to arbitrary fluids, which is, however, beyond the scope of the present paper.

The physical picture we put forward differs significantly from the picture developed in  \cite{Gangardt2009,schecter2012dynamics,schecter2012critical} for 1D systems. The method of \cite{Gangardt2009,schecter2012dynamics,schecter2012critical} is based on adiabatically following the total dispersion of the impurity-fluid system $\disp(p)$. Since $\disp(p)$ is periodic in one dimension, the authors of \cite{Gangardt2009,schecter2012dynamics,schecter2012critical} conclude that Bloch-like oscillations of the velocity of the impurity develop, provided $\disp(p)$ is a smooth function. This approach leaves no room for the SwO regime, in conflict with the results reported here and in Ref. \cite{Gamayun2014}. We note, however, that a key ingredient of the argument of Refs. \cite{Gangardt2009,schecter2012dynamics,schecter2012critical},  adiabaticity, as a rule can not be maintained for many-body gapless systems in the thermodynamic limit \cite{balian2007microphysics,polkovnikov2008breakdown,altland2008many,altland2009nonadiabaticity}. Although this issue has  triggered an active discussion \cite{schecter2014comment,Gamayun2014reply}, it is not resolved so far and requires further studies \cite{Burovski2015}.
Note  that the sawtooth oscillations in a 1D system  has also been discussed  in Ref. \cite{schecter2012critical}, but in in the limit of  strong force and only provided $\disp(p)$ has a cusp (see also a precursory work \cite{lamacraft2009dispersion}). These oscillations differ from those discussed here in amplitude and maximal velocity. We emphasize that smoothness of $\disp(p)$ plays no role in our arguments, in contrast to Refs.  \cite{schecter2012critical,lamacraft2009dispersion}.


{\em Summary and concluding remarks.}---
To summarise, we have studied two related settings. In the first setting a mobile impurity is injected with some initial velocity $v_0$ in a quantum fluid at zero temperature.  We have rigorously derived upper bounds \eqref{lower bound general} and \eqref{lower bound} on the difference between the initial and the asymptotic velocities of the impurity, $|\vv_0-\vv_\infty|$, valid for $|\vv_0|$ less than the mass-dependent generalized critical velocity~$v_c$.

These bounds imply that while the the velocity of the impurity can drop, it, generally speaking, does not drop to zero. This is consistent with the result of Ref. \cite{suzuki2014creation}: The impurity injected in the Bose-Einstein condensate creates a finite number of quasiparticles before relaxing to a steady state. On the other hand, our result disproves a suggestion of Refs. \cite{roberts2006force,roberts2009superfluids} (see also \cite{roberts2005casimir}) that perpetual motion of an impurity in a superfluid is nonexistent in thermodynamic limit due to the Casimir-like friction force.

We note that at any finite temperature $T$ the infinite-time velocity is most likely to vanish. The results \eqref{lower bound general} and \eqref{lower bound} remain relevant at low but nonzero temperatures if understood as bounds on the velocity at an intermediate timescale which is much less than the thermal relaxation timescale $\sim \hbar^7 m (v_s/k_{\rm B} T)^{2+2D} a^{-2 D}$, where $a$ is the scattering length \cite{klemens1955scattering,Baym1967,klemens1994thermal,CastroNeto1995}. For $D=3$ one gets relaxation timescale
$\sim 1\,{\mathrm s}
\left(\frac{m}{m_{\rm Rb}}\right)\left(\frac{v_s}{1\, {\rm mm}/{\rm s}}\right)^8 \left(\frac{T}{100 \, {\rm nK}}\right)^{-8} \left(\frac{a}{10 \, {\rm nm}}\right)^{-6}$
with $m_{\rm Rb}=85.47$ amu and other reference values relevant for ultracold atom experiments \cite{bloch2008many}.


In the second setting an impurity is pulled by a small constant force. We have demonstrated that, in general, two dynamical regimes can occur -- backscattering oscillations of the impurity velocity (BO) or  velocity saturation without oscillations (SwO). For fluids with $v_{c {\rm L}}=v_s$ SwO is the only possible regime. For fluids with $v_{c {\rm L}}<v_s$ SwO occurs for light impurities while BO occur for heavy impurities,  the two regimes being separated by a nonequilibrium quantum phase transition at some critical mass, see Table \ref{table} and inset in Fig. \ref{fig 1}.


Our treatment of the first problem is valid for any strength of impurity-fluid interaction, however the weaker is the interaction, the tighter are the bounds. Our treatment of the second problem is valid in the leading order of the weak coupling limit only. However, it is not necessarily the bare coupling which should be weak:  If one is able to find a renormalizing unitary transformation which takes into account the dressing of the impurity in a particular fluid and leads to a small effective coupling, this suffices to validate our treatment.

{\em Note added.}--- In a very recent paper \cite{castin2015vitesse} the concept of mass-dependent generalized critical velocity of a mobile impurity in a Fermi superfluid is studied in great detail. In particular, the nonanalyticity of $v_c$ as a function of mass is discussed.

\begin{acknowledgments}
{\em Acknowledgements.}
The author is grateful to V. Cheianov, O. Gamayun, E. Burovskiy, M. Zvonarev, G. Shlyapnikov, G. Pickett, P. McClintock, V. Tsepelin,  A. Fedorov and M. Schecter for fruitful discussions. The present work was supported by the ERC grant 279738-NEDFOQ.
\end{acknowledgments}

\bibliography{D:/Work/QM/Bibs/critical_velocity,D:/Work/QM/Bibs/1D,D:/Work/QM/Bibs/He,D:/Work/QM/Bibs/LZ_and_adiabaticity,D:/Work/QM/Bibs/phonon_scattering}

\end{document}